\documentclass[graybox]{svmult}

\usepackage{mathptmx}       
\usepackage{helvet}         
\usepackage{courier}        
\usepackage{type1cm}        
\usepackage{makeidx}         
\usepackage{graphicx}        
\usepackage{multicol}        
\usepackage[bottom]{footmisc}

\makeindex             
             
\usepackage{amsmath,amssymb}
\usepackage[misc]{ifsym}
\usepackage{url,hyperref}
\usepackage{subfig}
\newcommand\dlangle{\langle\!\langle}
\newcommand\drangle{\rangle\!\rangle}

\begin{document}

\title*{A Parametric Study of Mixing in a Granular Flow a Bi-Axial Spherical Tumbler}
\titlerunning{Parametric Study of Granular Mixing in a Spherical Tumbler}
\author{Ivan C.\ Christov, Richard M.\ Lueptow, Julio M.\ Ottino and Rob Sturman}
\authorrunning{I.~C.~Christov, R.~M.~Lueptow, J.~M.~Ottino and R.~Sturman}
\institute{Ivan C. Christov (\Letter) \at Theoretical Division and Center for Nonlinear Studies, Los Alamos National Laboratory, Los Alamos, NM 87545, USA, \email{christov@alum.mit.edu}
\and Richard M. Lueptow \at Department of Mechanical Engineering and The Northwestern Institute on Complex Systems (NICO), Northwestern University, Evanston, IL 60208, USA, \email{r-lueptow@northwestern.edu}
\and Julio M. Ottino \at Department of Chemical and Biological Engineering, Department of Mechanical Engineering and The Northwestern Institute on Complex Systems (NICO), Northwestern University, Evanston, IL 60208, USA, \email{jm-ottino@northwestern.edu}
\and Rob Sturman \at Department of Applied Mathematics, University of Leeds, Leeds LS2 9JT, UK, \email{r.sturman@maths.leeds.ac.uk}%
}
%
%
\maketitle

\abstract{We report on a computational parameter space study of mixing protocols for a half-full bi-axial spherical granular tumbler. The quality of mixing is quantified via the intensity of segregation (concentration variance) and computed as a function of three system parameters: angles of rotation about each tumbler axis and the flowing layer depth. Only the symmetric case is considered in which the flowing layer depth is the same for each rotation. We also consider the dependence on $\bar{R}$, which parametrizes the concentric spheroids (``shells'') that comprise the volume of the tumbler. The intensity of segregation is computed over 100 periods of the mixing protocol for each choice of parameters. Each curve is classified via a time constant, $\tau$, and an asymptotic mixing value, $bias$. We find that most choices of angles and most shells throughout the tumbler volume mix well, with mixing near the center of the tumbler being consistently faster (small $\tau$) and more complete (small $bias$). We conclude with examples and discussion of the pathological mixing behaviors of the outliers in the so-called $\tau$-$bias$ scatterplots.}

\section{Introduction}
\label{sec:intro}

Most granular flows encountered in nature and industry are three-dimensional (3D). However, the mathematical modeling and analysis, and the experimental and computational visualization of 3D flows remains a challenge. Yet, 3D granular flows reveal many features of complex systems far from equilibrium. Rotating drums, also known as tumblers, are one of six fundamental systems in which dense granular flows are studied \cite{midi04} and are a common device in the pharmaceutical and other powder processing industries \cite{mm09}. Here, we consider a simple, but realistic, analytically-tractable 3D granular flow termed the blinking spherical tumbler flow because of its analogy to an early example of 2D chaotic mixing \cite{o89}: the blinking vortex flow \cite{a84}.

A quasi-3D granular flow, specifically weak rocking motions of a rotating cylindrical container,  appears to be one of the first experimentally and numerically studied examples \cite{WightmanSim}. More recently, spherical tumblers were analyzed experimentally and computationally \cite{go03,jlosw10,mlo07,zduol12}. Through a dynamical systems framework, it has become possible to gain a deeper understanding of how mixing and segregation manifest themselves in these 3D systems \cite{mlo07,smow08,clos15}. Even so, chaotic transport and mixing in 3D flows remains relatively unexplored, especially given the great variety of new (i.e., distinct from those in 2D) behaviors possible \cite{wig-fof}.

In the present work, we are interested in \emph{quantifying} mixing in a 3D chaotic granular flow. Although many recent studies have focused on understanding the kinematic structures created by 3D chaotic flows and maps, only a few studies quantify mixing in a \emph{global} way for \emph{genuinely} 3D flows: e.g., via numerical simulations in a cavity flow with translating side-walls \cite{agpvm00,gakpm01} and in an ``ABC'' flow \cite{scuol13}, and via experiments in a container with counter-rotating lids \cite{ls06} and for granular flows in blade and bin blenders \cite{psfe04,mm09,rgk10}. Given the uniquely distinct dynamics of granular flow in a bi-axial spherical tumbler, here we seek to connect the ``degree of mixing'' to the independent parameters of this system and to also understand how mixing varies across the volume of a half-full tumbler. To this end, we present a parametric study of mixing in the so-called symmetric case of a half-full blinking spherical tumbler using the technique of $\tau$-$bias$ scatterplots \cite{mcwig}.

\section{Continuum Model of Granular Flow in a Spherical Tumbler}
\label{sec:3d_const_ang}

\begin{figure}
\sidecaption
\centering
\includegraphics[width=0.625\textwidth]{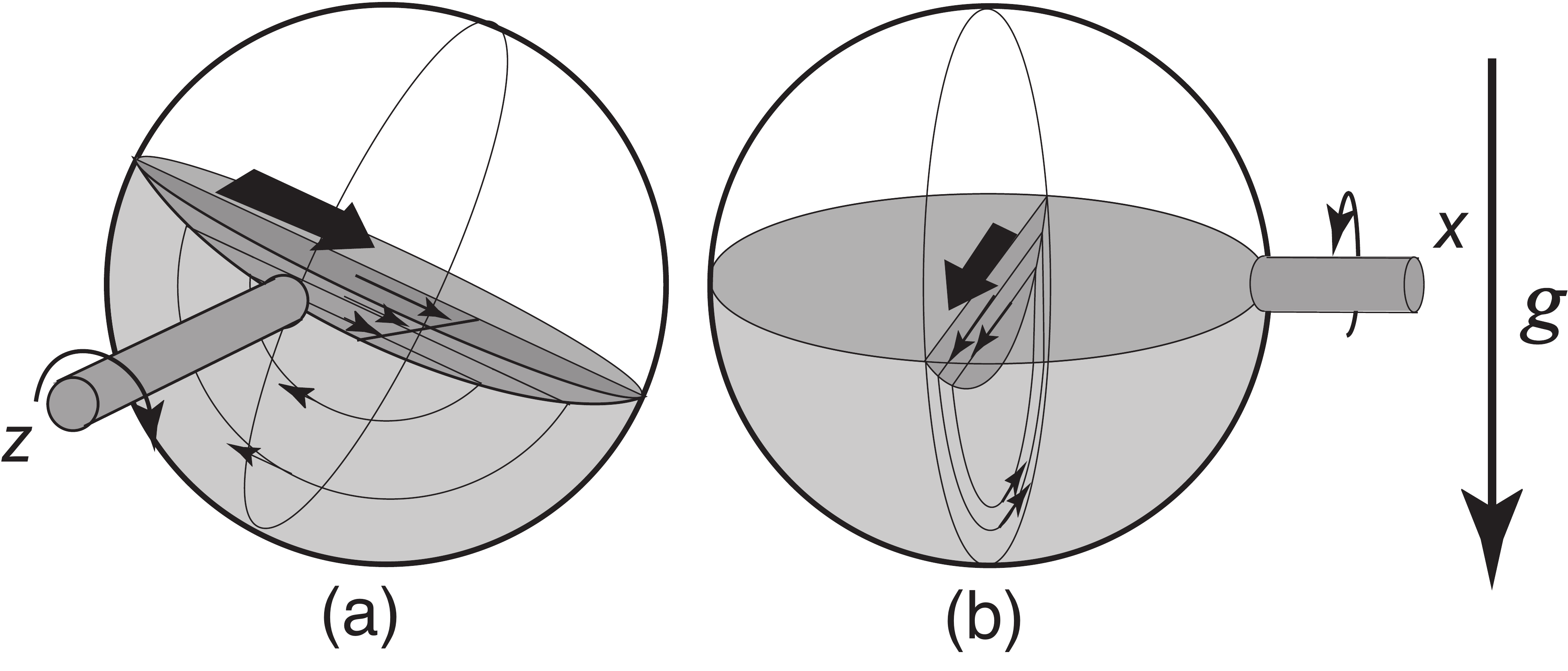}
\caption{Schematic of a half-full blinking spherical tumbler. Streamlines are shown in the cross-sections for rotation around the (a) $z$- and (b) $x$-axis. Flow in each cross-section consists of solid body rotation in the bulk, followed by a rapid cascade down a thin flowing layer at the free surface. Reproduced with permission from \cite{jlosw10}.}
\label{fig:biaxsphere}
\end{figure}

In this section, we summarize only the necessary details of the continuum model (see, e.g., 
\cite{mlo07,c11} for derivations) for the kinematics of granular flow in the continuous-flow (rolling) regime of a half-full spherical tumbler rotated about two orthogonal axes (for the present purposes, the $z$- and $x$-axes as shown in Fig.~\ref{fig:biaxsphere}).


Without loss of generality, we take the first rotation of the half-full spherical tumbler of radius $R$ to be clockwise about the $z$-axis at a rate of $\omega_z(>0)$ for a duration $T_z$. We employ a Cartesian coordinate system with its origin at the center of the sphere, the initial streamwise direction is along the $x$-axis, and the transverse direction is along the $z$-axis. The streamwise velocity $v_x$ varies linearly with the depth $y$ in the flowing layer so that the shear rate $\dot{\gamma_z} = \partial v_x/\partial y$ is constant.\footnote{In an experiment, the flowing layer is at an angle with respect to the horizontal, while, in the model, the coordinate system is rotated backwards by this fixed angle. Also, we assume that the effects of side-walls and transport in the transverse direction are negligible.}

We make $x$, $y$, $z$, $L$, $\delta$ dimensionless by rescaling by $1/R$ and $t$ is made dimensionless by rescaling by $\omega_z$ (for rotation about the $z$-axis) or $\omega_x$ (for rotation about the $x$-axis).
The velocity field in the bulk (bed of solid body rotation) is then 
\begin{equation}
\dot{x} = y,\qquad  \dot{y} = -x,\qquad \dot{z} = 0;\qquad y\le-\delta_z(x,z).
\label{eqs:u_rotz_b}
\end{equation}
The boundary of the flowing layer \cite{mlo07,smow08,clos15} is the surface
\begin{equation}
\delta_z(x,z) = \delta_{0}(z) \sqrt{1-{x^2}/{L(z)^2}} = \epsilon_z\sqrt{1 - x^2 - z^2},
\label{eq:delta_z}
\end{equation}
where $\delta_{0}(z) = L(z) \sqrt{{\omega_z}/{\dot{\gamma_z}}}$ and $L(z) = \sqrt{1-z^2}$ are  the local (dimensionless) depth and half-length of the flowing layer, respectively. Naturally, the definition of $\epsilon_z := \delta_{0}(0)/L(0) \equiv \sqrt{{\omega_z}/{\dot{\gamma_z}}}$ as the maximal depth at $x=z=0$ follows since $L(0)=1$.
Meanwhile, the velocity field in the flowing layer \cite{mlo07,smow08,clos15} can be written as
\begin{equation}
\dot{x} = \left[\delta_z(x,z)+y\right]/\epsilon_z^{2},\qquad
\dot{y} = {x\,y}/{\delta_z(x,z)},\qquad
\dot{z} = 0;\qquad y>-\delta_z(x,z).
\label{eqs:u_rotz_fl}
\end{equation}
In dimensionless units, the tumbler is rotated about the $z$-axis by the angle $\theta_z = \omega_z T_z$.



The blinking spherical tumbler protocol alternates between rotations about two orthogonal axes (Fig.~\ref{fig:biaxsphere}). Once the rotation about the $z$-axis is complete, a rotation about the $x$-axis at constant rate $\omega_x$ for a duration $T_x$ follows,\footnote{In an experiment, several intermediate steps are performed between the rotations \cite{mlo07}, however, as far as the mathematical analysis is concerned, the rotations are completely independent.} with the streamwise direction now along the $z$-axis and Eqs.~\eqref{eqs:u_rotz_b}--\eqref{eqs:u_rotz_fl} still holding after formally exchanging ``$x$'' and ``$z$.''  
Specifically, the interface between the flowing layer and the bulk for rotation about the $x$-axis is given by
$\delta_x(x,z) = \epsilon_x\sqrt{1-x^2-z^2}$,
where $\epsilon_x = \sqrt{\omega_x/\dot\gamma_x}$. 

Note that if $\epsilon_z = \epsilon_x$, then $\delta_z(x,z) \equiv \delta_x(x,z)$. We refer to this as the \emph{symmetric case}. Naturally, $\epsilon_z \ne \epsilon_x$ will be termed the \emph{non-symmetric case}. It is important to note that different values for $\epsilon_z$ and $\epsilon_x$ arise in physical systems due to different rotation rates $\omega_z$ and $\omega_x$ \cite{pakcol12}. Thus, the symmetric case occurs for identical rotation rates about the two axes, while the non-symmetric case arises from different rotation rates. 

%

\section{Parametric Study of Mixing in the Symmetric Case}
\label{sec:sym_mix}


While previous studies \cite{mlo07,smow08,clos15} have focused on the qualitative structure of the template for chaotic transport in the blinking spherical tumbler, here we study the ``quality of mixing'' quantitatively by measuring the decay of the variance of the concentration of a species. For the remainder of the discussion, we restrict the analysis to the symmetric case (i.e., $\epsilon_z = \epsilon_x$) of the half-full blinking spherical tumbler, and we fix the initial condition (IC) such that tracer particles within the quarter-sphere 
$\mathfrak{Q} = \big\{ (x,y,z) \in \mathbb{R}^3 \;|\; x^2 + y^2 + z^2 \le 1,\; -1\le y \le 0,\; 0 \le x \le 1\big\}$
are colored gray at $n=0$, while those in the remainder of the filled volume (i.e., $\{ (x,y,z) \in \mathbb{R}^3 \;|\; x^2 + y^2 + z^2 \le 1,\; -1\le y \le 0,\; -1\le x < 0\}$) are black. 
From the assumption of $\epsilon_z=\epsilon_x$, as shown in \cite{smow08,clos15}, it follows that trajectories of these tracer particles are restricted to 2D invariant surfaces, which are hemispherical in the bulk, determined by their ICs \cite{clos15}. Therefore, it is convenient to introduce the ``artificial'' parameter, termed the \emph{shell radius}, $\bar{R} \in (\epsilon_z,1)$ that denotes the (constant) radius from the origin of the hemispherical bulk portion of the 2D invariant surfaces upon which tracer particles are restricted. Since the dynamics depend strongly upon which hemispherical shell is considered \cite{clos15}, the shell radius $\bar{R}$ is a useful parameter.

Given the non-dimensionalization introduced in Section~\ref{sec:3d_const_ang}, it is clear there are four independent parameter in our model of the half-full blinking spherical tumbler, namely $\epsilon_z$, $\epsilon_x$, $\theta_z$ and $\theta_x$. Indeed, in a quasi-2D experiment, there are only four quantities that vary independently once the tumbler and granular material are chosen \cite{pakcol12}. Specifically, fixing the rotation rate in the laboratory yields unique values for the shear rate and maximal depth of the flowing layer, though the precise functional relationships remains a topic of research (see, e.g., \cite{pakcol12} and the references therein). 
In our model, we set the flowing layer depth, which is thus equivalent to setting the rotation rate in an experiment. Additionally, in our model there may be different angles of rotation about each axis but the flowing layer depths for rotation about each axis are equal ($\epsilon_x=\epsilon_x$) because we have limited the analysis to the symmetric case. Therefore, including $\bar{R}$, there are four independent parameters. 

With this in mind, we can ``bin'' (see \cite[Appendix B]{c11} for details) the part of the invariant surface in the bulk into $N\times N$ surface elements, each of area $A_i$. 
For the sake of simplicity we construct bins with equal areas, i.e., $A_i\equiv A$ $\forall i$. In each, we uniformly distribute $M_b\times M_b$ tracer particles throughout the area. Then, these particles are advected for $p$ flow periods, where the flow period is the total time required to complete a rotation about each of the two axes (i.e., an iteration of the blinking protocol). Finally, the concentration $c_i$ of gray (equivalently, black) tracer particles can be computed in each of the original bins as
$c_i = {\#_{\mathrm{g},i}}/({\#_{\mathrm{g},i} + \#_{\mathrm{b},i}})$,
where $\#_{\mathrm{g},i}$ and $\#_{\mathrm{b},i}$ denote the number of gray and black particles in the $i$th bin, respectively. Following Danckwerts~\cite{d52}, consider the first moment (mean) of this distribution:
\begin{equation}
\langle c \rangle = \frac{\sum_{i=1}^{N^2}c_iA_i}{\sum_{i=1}^{N^2} A_i} = \frac{1}{N^2}\sum_{i=1}^{N^2}c_i,
\end{equation}
where the last equality follows from the fact that $A_i\equiv A$ $\forall i$. Then, the second moment (variance) of the distribution of concentrations is given by
\begin{equation}
\dlangle c\drangle \equiv \langle (c - \langle c\rangle)^2 \rangle = \frac{\sum_{i=1}^{N^2}(c_i-\langle c \rangle)^2 A_i}{\sum_{i=1}^{N^2} A_i} = \frac{1}{N^2}\sum_{i=1}^{N^2} (c_i-\langle c \rangle)^2.
\end{equation}
It is convenient to normalize $\dlangle c\drangle$ by introducing the \emph{intensity of segregation} \cite{d52}:
\begin{equation}
\mathcal{I} = \frac{\dlangle c\drangle}{\langle c\rangle (1 - \langle c\rangle)},
\end{equation}
where now $0\le \mathcal{I}\le 1$ with $\mathcal{I}=0$ and $1$ corresponding to perfect mixing (i.e., $c = \langle c\rangle$ everywhere) and perfect segregation (i.e., $c = 0$ or $1$ for any given surface element in the domain), respectively.

For the simulations presented below, we divide the bulk (hemispherical) portion of the 2D invariant surface on which the dynamics are restricted into $N=14$ longitudinal and $N=14$ latitudinal bins with $M_p\times M_p=10^2$ uniformly distributed tracers in each ($\Rightarrow 19,600$ tracers total). 
These are advected for a total for $p=200$ flow periods using the exact solution for rotations about the $z$- and $x$-axis given in \cite{c11,clos15}. The intensity of segregation $\mathcal{I}(n)$ is tracked as a function of the period $n=1,\hdots,p$. Following \cite{mcwig}, using {\sc Matlab}'s Curve Fitting Toolbox we find $\tau$ and $bias$ such that the fit based on the following ansatz is ``best'' in a least-squares sense:\footnote{In Figs.~\ref{fig:sym_mix_ex_thick},\ref{fig:sym_mix_ex_thick2}--\ref{fig:sym_mix_ex_thin2} below, the quality of this fit is given as two numbers in brackets in the legend of the rightmost plot. The first number is the coefficient of determination $R^2$; the second number is the root-mean-squared error in the fit.}
\begin{equation}
\mathcal{I}(n) \sim \left(\mathcal{I}(0) - bias\right)\mathrm{e}^{-n/\tau} + bias,
\label{eq:fit}
\end{equation}
where $\mathcal{I}(0) = 1$ for the chosen completely segregated IC. 
Note that $\tau$ is dimensionless and has the meaning of ``number of periods,'' while $bias$ allows for cases where $\mathcal{I} \nrightarrow 0$ as $n \to \infty$ (see, e.g., the discussion in \cite{l54}). 

An example is shown in Fig.~\ref{fig:sym_mix_ex_thick} for two different values of $\theta_z$ and $\theta_x$ and at two shell radii $\bar{R}$. The left images show the mixing of gray and black 100 iterations after starting completely segregated. The right graphs show the decay of $\mathcal{I}$. In both cases, $\mathcal{I}$ starts at $1$ due to the completely segregated IC. The case in Fig.~\ref{fig:sym_mix_ex_thick}(a) mixes to some extent but not completely after 100 iterations, resulting in a relative large $bias$ of $0.15$. The case in Fig.~\ref{fig:sym_mix_ex_thick}(b) mixes quickly and completely resulting in a smaller $\tau$ than the case in Fig.~\ref{fig:sym_mix_ex_thick}(a) and a $bias$ of almost $0$.

\begin{figure}
\centering
\subfloat[$\theta_z = 30^\circ$, $\theta_x = 5^\circ$ and $\bar{R} = 0.9$]{\includegraphics[width=\textwidth]{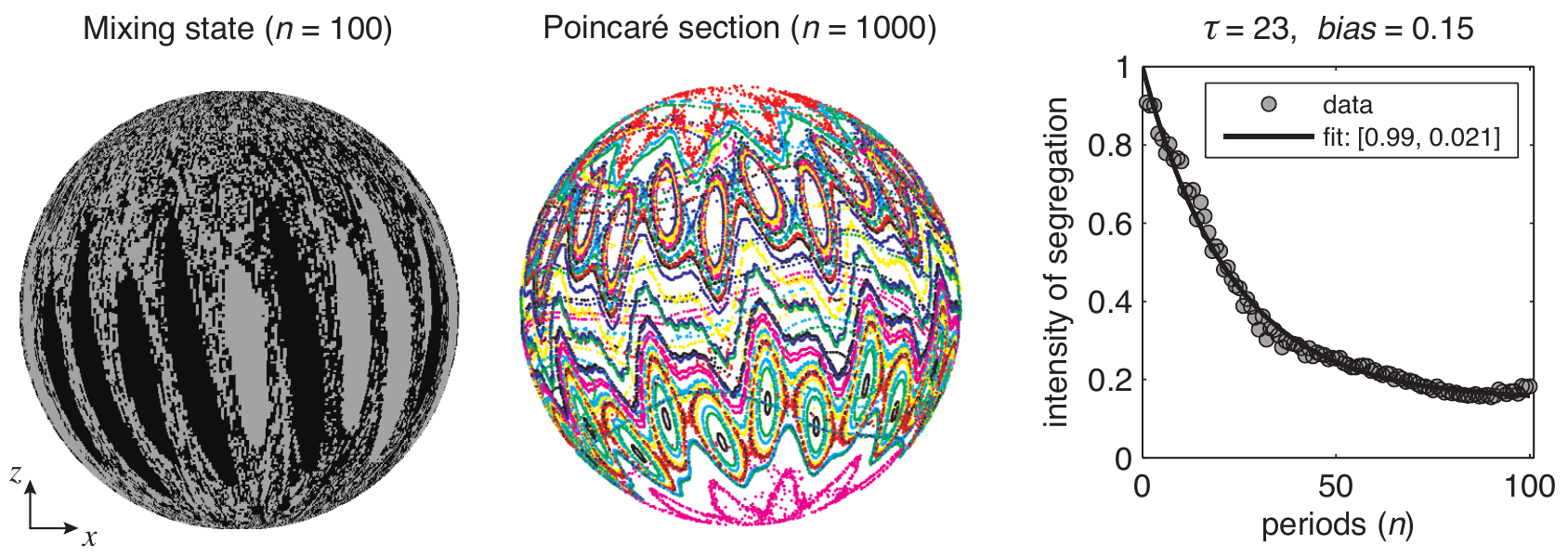}}\\
\subfloat[$\theta_z = 45^\circ$, $\theta_x = 65^\circ$ and $\bar{R} = 0.5$]{\includegraphics[width=\textwidth]{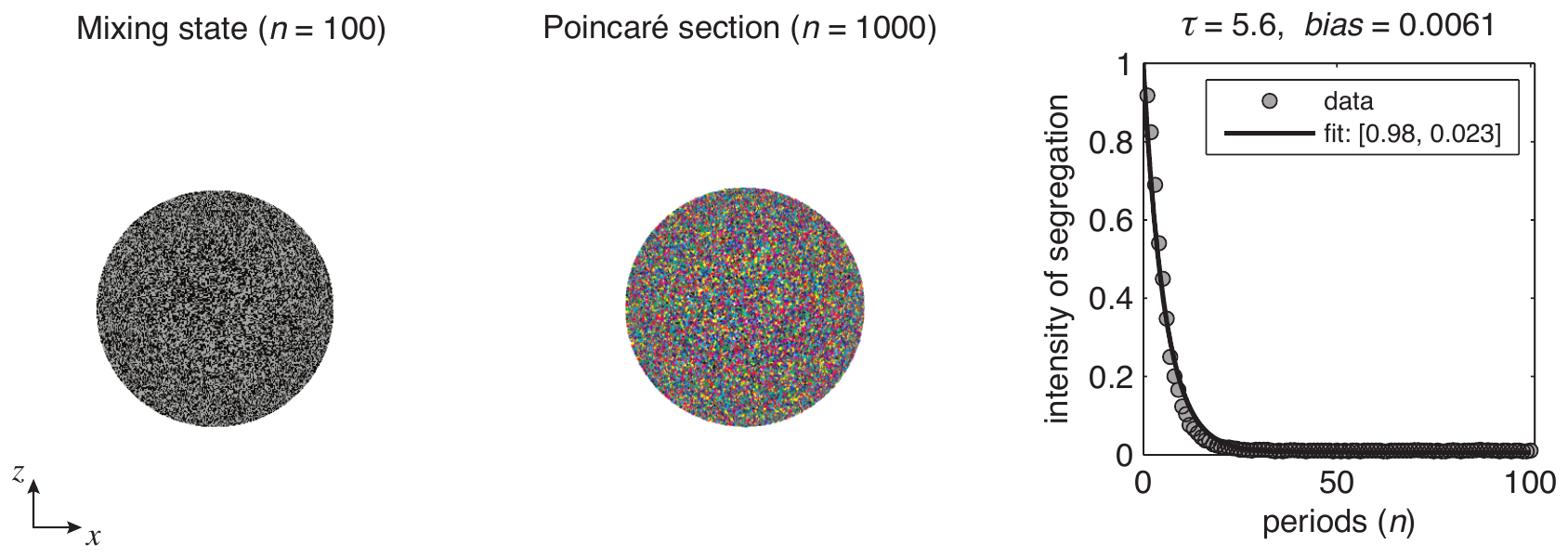}}
\caption{Examples of mixing behavior in the half-full blinking spherical tumbler with symmetric typical-thickness flowing layers: $\epsilon_z = \epsilon_x = 0.15$.}
\label{fig:sym_mix_ex_thick}
\end{figure}

Therefore, we have two numbers, i.e., $\tau$ and $bias$, that measure the ``quality of mixing'' for each numerical mixing experiment---$\tau$ is a time constant measuring how quickly the mixture is homogenized, while $bias$ is a measure of how thoroughly the asymptotic state is mixed. Large $bias$ corresponds to incomplete mixing and suggests the presence of large unmixed Kolmogorov--Arnold--Moser (KAM) ``islands.'' Meanwhile, a large $\tau$ corresponds to slow mixing, say on a surface with large area with sub-optimal choice of rotation angles. Finally, a low $\tau$ and a low $bias$ correspond to ``good'' mixing as the homogenization of the segregated IC is fast and thorough. Next, we study how $\tau$ and $bias$ vary with the parameters of the model.

\begin{figure}
\centering
\includegraphics[width=\textwidth]{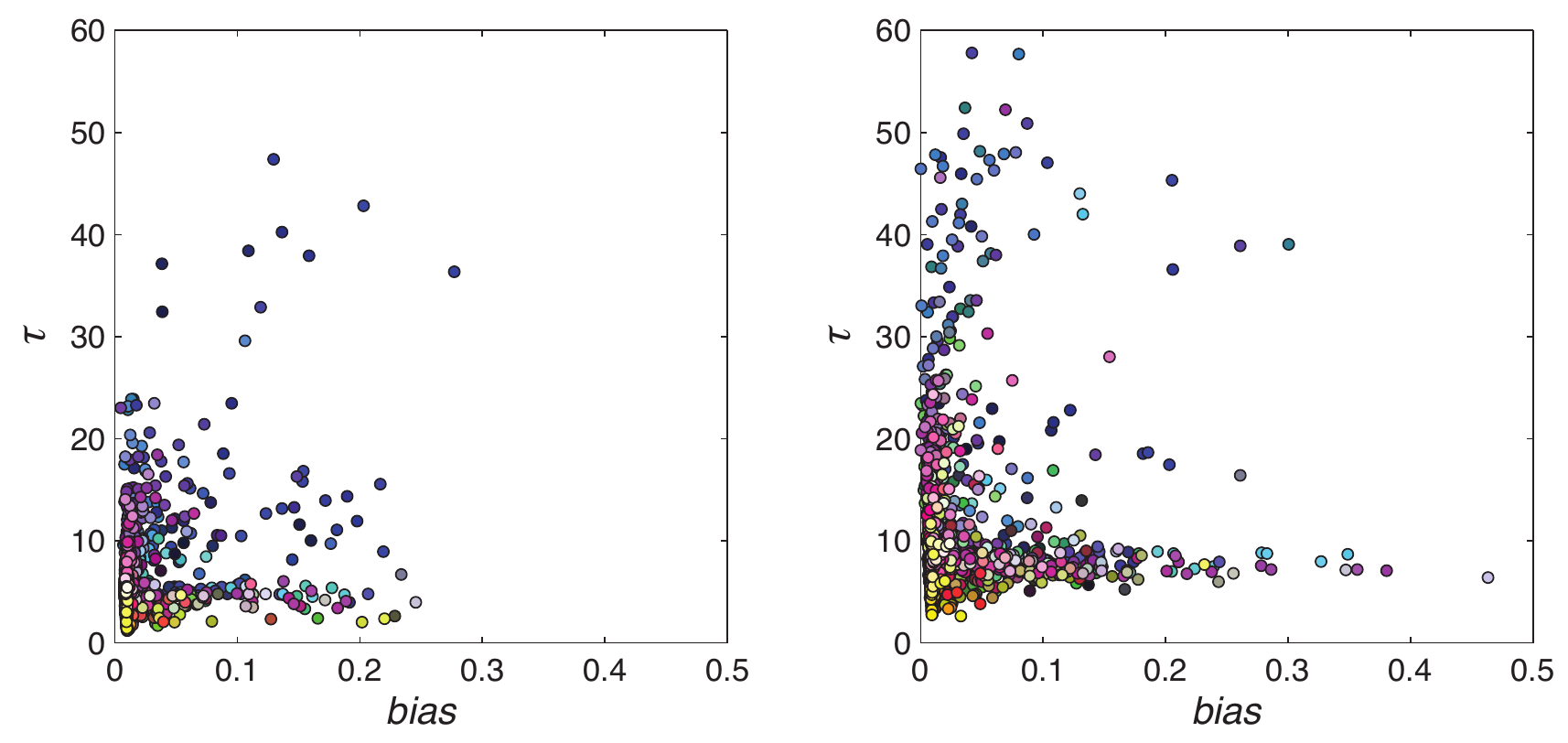}
\caption{$\tau$-$bias$ scatterplots for $\epsilon_z = \epsilon_x = 0.15$ (left) and $\epsilon_z = \epsilon_x = 0.05$ (right). $\mathcal{I}$ is computed on hemispherical shells with $\bar{R}$ as a parameter. Data points are color-coded based on the values of $\theta_z$, $\theta_x$ and $\bar{R}$ as described in the text.}
\label{fig:tb_scatter}
\end{figure}

The most natural parameters to vary are the angles of rotation about each axis, $\theta_z$ and $\theta_x$, and the shell radius $\bar{R}$. Additionally, $\epsilon_z$ ($\equiv\epsilon_x$ in the symmetric case) can be varied, corresponding to a different type of granular material in the tumbler or a different rotation rate. To ascertain the effect of the variation of $\theta_z$, $\theta_x$ and $\bar{R}$, in Fig.~\ref{fig:tb_scatter} we show a scatterplot of $\tau$ versus $bias$ for independent choices of $\theta_z$ and $\theta_x$ between $5^\circ$ and $355^\circ$ in increments of $25^\circ$ and 10 equispaced values of $\bar{R}$ between $\epsilon_z+0.1$ and $0.95$. 
Additionally, two values of $\epsilon_z$ are considered: (a) $\epsilon_z=0.15$, a value realized for a dry granular system of 3 mm chrome steel beads in a tumbler of 28 cm diameter rotating at 0.168 rad/s, and (b) $\epsilon_z = 0.05$, a thinner flowing layer observed for 1.16 mm steel beads in this same tumbler rotating at 0.052 rad/s \cite{jol04}.\footnote{Though these numbers are for a quasi-2D circular tumbler, they are relevant here too thanks to the geometric similarity assumption used to construct the 3D continuum model. That is to say, if we suppose the quasi-2D tumbler of Jain et al.~\cite{jol04} is the $x=0$ (or $z=0$) cut through the 3D spherical tumbler considered here, then $\epsilon_z$ (or $\epsilon_x$) here is precisely $\delta_0/R$ in \cite{jol04}.} 

Each data point in Fig.~\ref{fig:tb_scatter} is color-coded as follows: it has an red-green-blue (RGB) additive color intensity with \%R given by $\theta_z/(2\pi)$, \%G given by $\theta_x/(2\pi)$ and \%B given by $\bar{R}$. Hence, data points in the blue spectrum correspond to angles of rotation close to $5^\circ$ and shell radii $\bar{R}$ close to 1. Similarly, data points that appear orange correspond to angles of rotation close to $355^\circ$ but shell radii close to $\epsilon_z+0.1$. Likewise, white corresponds to both $\theta_z$ and $\theta_x$ near $355^\circ$ with $\bar{R}$ near 1, and so forth.

First, consider the black-blue-purple range of colors of data points, which  correspond to $\theta_z$ and $\theta_x$ both near $5^\circ$ for a variety of $\bar{R}$. It is immediately clear that these represent the majority of the ``outliers'' in the scatterplots in Fig.~\ref{fig:tb_scatter}. To see why, recall that our IC 
is symmetric about $z=0$ and anti-symmetric about $x=0$. In addition, $\theta_z$ is the angle of rotation about the $z$-axis, which is the first axis of rotation in this protocol. It follows that for values of $\theta_z$ close to $5^\circ$, the IC is barely altered by the rotation about the first axis. Then, since the angle of rotation about the second axis is small ($\theta_x$ near $5^\circ$ for these data points), the almost-$z$-symmetric state that the mixture is left in leads to very little re-arrangement of the material during the second rotation. Figures~\ref{fig:sym_mix_ex_thick}(a) and \ref{fig:sym_mix_ex_thin}(a) illustrates a typical such scenario for $\epsilon_z(=\epsilon_x)=0.15$ and $\epsilon_z(=\epsilon_x)=0.15$, respectively. We term it ``lots of cutting but no shuffling,'' using the terminology introduced in \cite{smow08,jlosw10}, due to the pattern of thin filaments that are barely re-oriented from each other that emerges in the mixture after 100 periods. Not surprisingly, the corresponding Poincar\'e sections in Figs.~\ref{fig:sym_mix_ex_thick}(a) and \ref{fig:sym_mix_ex_thin}(a) show a great deal of regularity, especially in Fig.~\ref{fig:sym_mix_ex_thick}(a), rather than chaos. As $\theta_z$ becomes larger, the distribution of material is ``less symmetric'' about the $z$ when the rotation about the $x$ begins, so the $(\tau,bias)$ pairs, which are now in the pink color spectrum, are no longer on the outer edges of the scatterplot.

\begin{figure}
\centering
\subfloat[$\theta_z = \theta_x = 180^\circ$ and $\bar{R} = 0.62$]{\includegraphics[width=\textwidth]{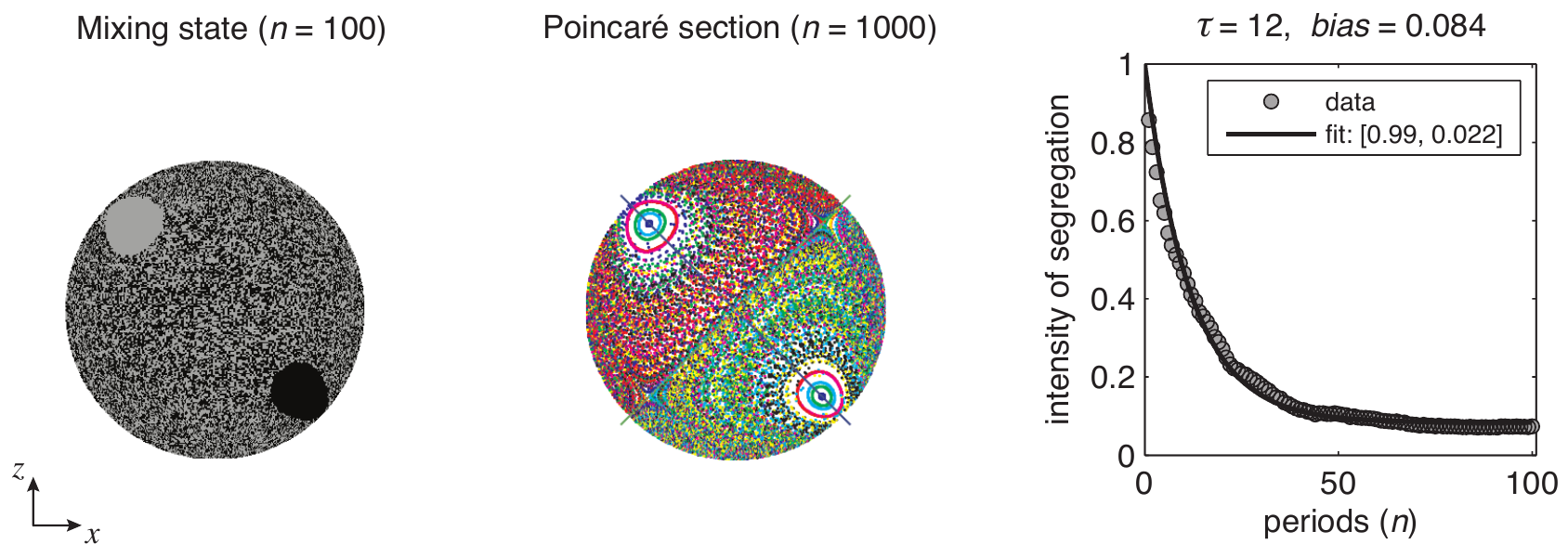}}\\
\subfloat[$\theta_z = 355^\circ$, $\theta_x = 180^\circ$ and $\bar{R} = 0.88$]{\includegraphics[width=\textwidth]{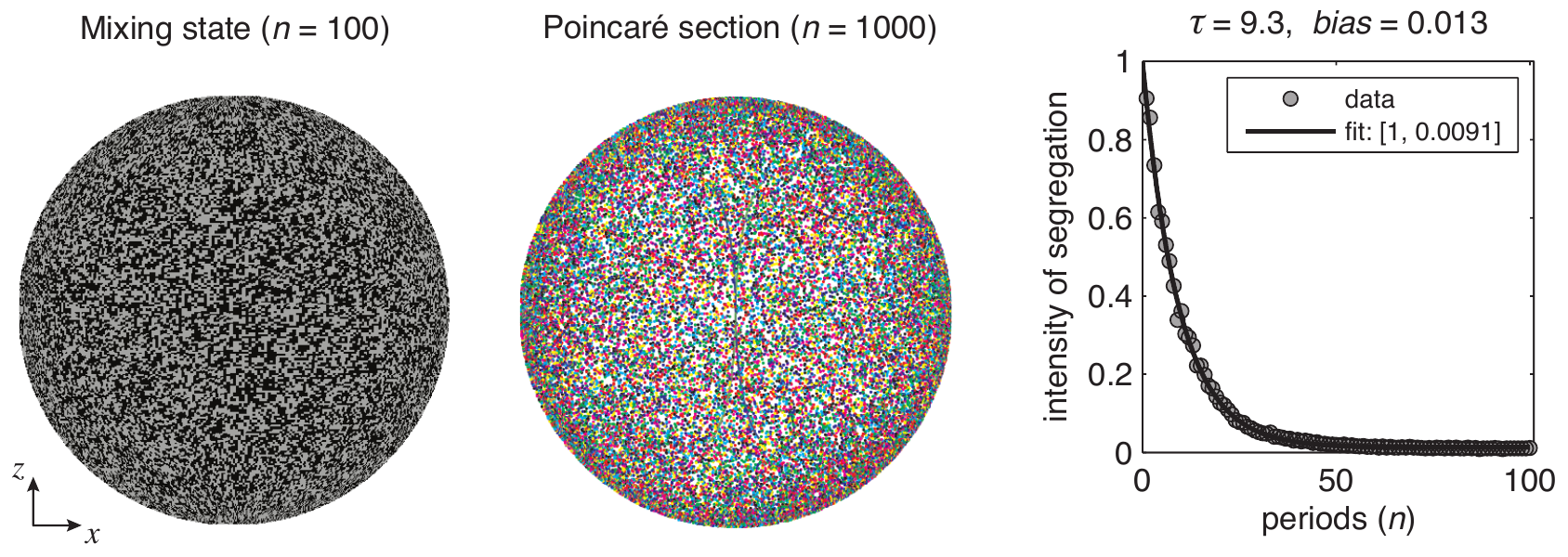}}
\caption{Further examples of mixing behavior in the half-full blinking spherical tumbler with symmetric typical-thickness flowing layers: $\epsilon_z = \epsilon_x = 0.15$.}
\label{fig:sym_mix_ex_thick2}
\end{figure}

Next, we observe the large cluster of data points, situated near $(\tau,bias) = (0,0)$ in the scatterplots, that have mostly colors in the white-yellow-orange-red spectrum. From the color-coding convention, these correspond to choices of $\theta_z$ near $355^\circ$, and choices of $\theta_x$ that increase proportionately with $\bar{R}$. Notice this avoids the ``lots of cutting and no shuffling'' scenario previously described. 
Figures~\ref{fig:sym_mix_ex_thick}(b) and \ref{fig:sym_mix_ex_thin}(b) show such cases for $\epsilon_z(=\epsilon_x)=0.15$ and $\epsilon_z(=\epsilon_x)=0.05$, respectively. It appears the dynamics on the 2D invariant surfaces on which the dynamics are restricted for these sets of parameters are fully chaotic, i.e., mixing occurs and no KAM islands or other barriers to transport can be distinguished. This is further supported by considering the Poincar\'e sections in Fig.~\ref{fig:sym_mix_ex_thick}(b) and \ref{fig:sym_mix_ex_thin}(b), which exhibit no regularity of any kind.

\begin{figure}
\centering
\subfloat[$\theta_z = 30^\circ$, $\theta_x = 180^\circ$ and $\bar{R} = 0.71$]{\includegraphics[width=\textwidth]{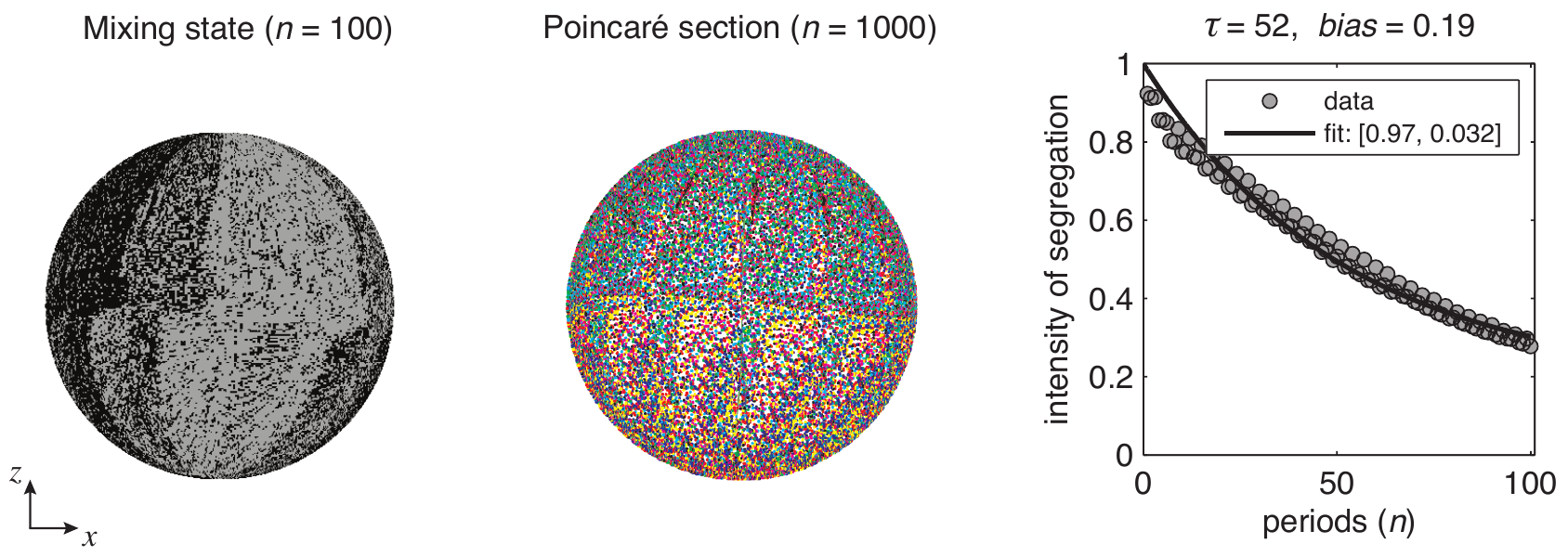}}\\
\subfloat[$\theta_z = 305^\circ$, $\theta_x = 55^\circ$ and $\bar{R} = 0.23$]{\includegraphics[width=\textwidth]{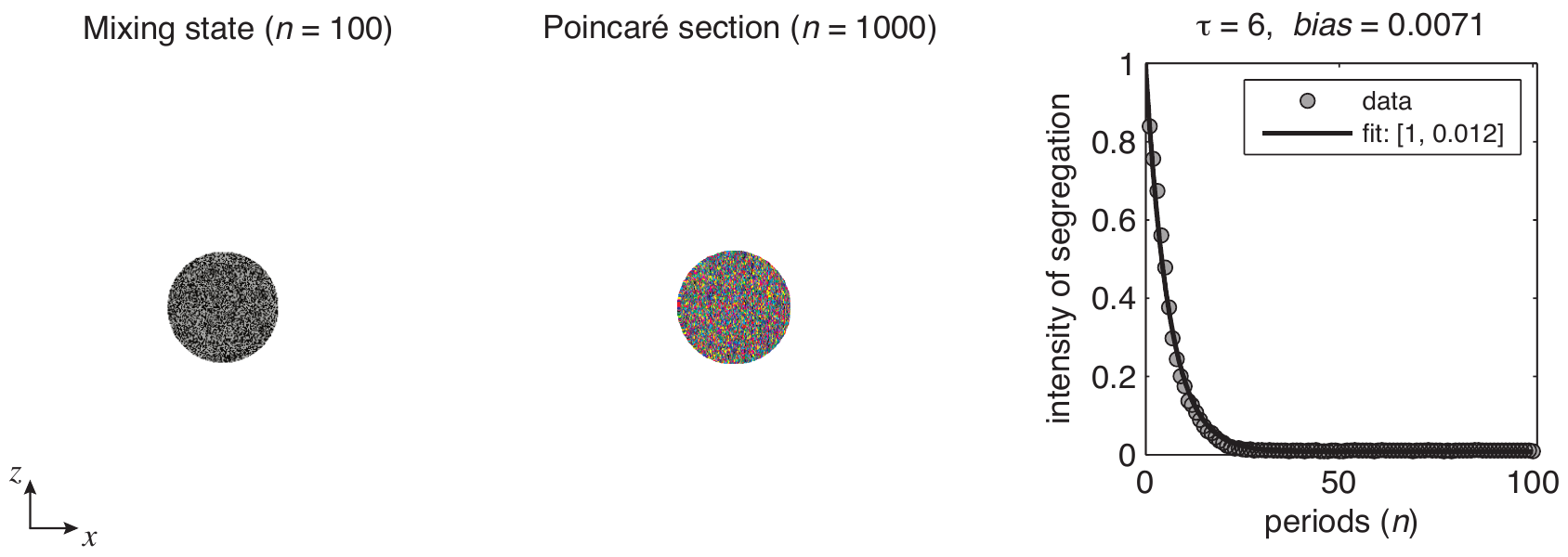}}
\caption{Examples of mixing behavior in the half-full blinking spherical tumbler with symmetric thin flowing layers: $\epsilon_z = \epsilon_x = 0.05$.}
\label{fig:sym_mix_ex_thin}
\end{figure}

As mentioned earlier, low-period islands are barriers to transport and they lead to incomplete mixing \cite{o89}. An example of period-one KAM islands can be found in Fig.~\ref{fig:sym_mix_ex_thick2}(a), while and example of period-two islands is presented in Fig.~\ref{fig:sym_mix_ex_thin2}(a). Though $bias$ is significant for both of these cases, $\tau$ is not very large, showing that a significant ``chaotic sea'' exists between the islands and the material there is quickly homogenized. This clearly seen in both the mixing patterns and the Poincar\'e sections in Figs.~\ref{fig:sym_mix_ex_thick2}(a) and \ref{fig:sym_mix_ex_thin2}(a), with the islands from the Poincar\'e sections clearly corresponding to unmixed regions of the same shape.

Independent of low-period structures, some protocols mix very slowly. Examples of such protocols that do eventually homogenize most of the mixture are show in Fig.~\ref{fig:sym_mix_ex_thick2}(b) and \ref{fig:sym_mix_ex_thin2}(b). This is not due to 
KAM islands because, as the Poincar\'e sections corresponding to these figures show, there are none on these shells. It is simply that these protocols do not re-orient material effectively due to the choice of rotation angles (i.e., $\theta_z \approx 360^\circ$). Consequently, mixing is ``suboptimal.''

\begin{figure}
\centering
\subfloat[$\theta_z = \theta_x = 85^\circ$ and $\bar{R} = 0.4$]{\includegraphics[width=\textwidth]{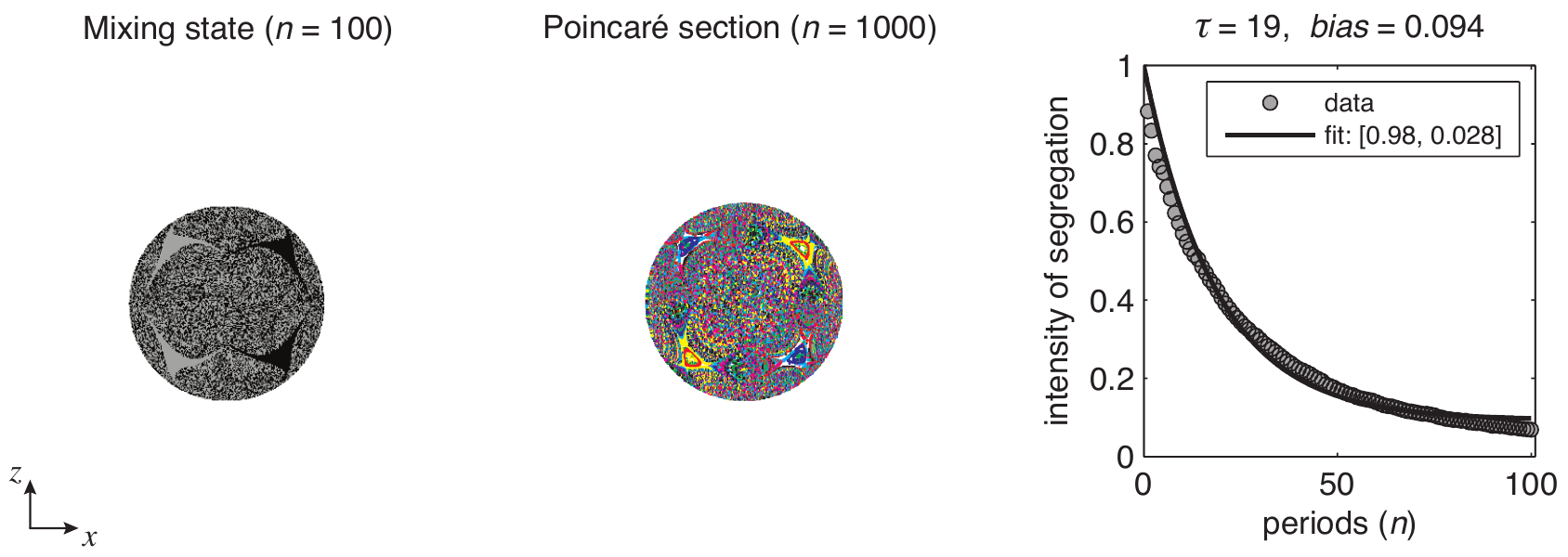}}\\
\subfloat[$\theta_z = 355^\circ$, $\theta_x = 205^\circ$ and $\bar{R} = 0.88$]{\includegraphics[width=\textwidth]{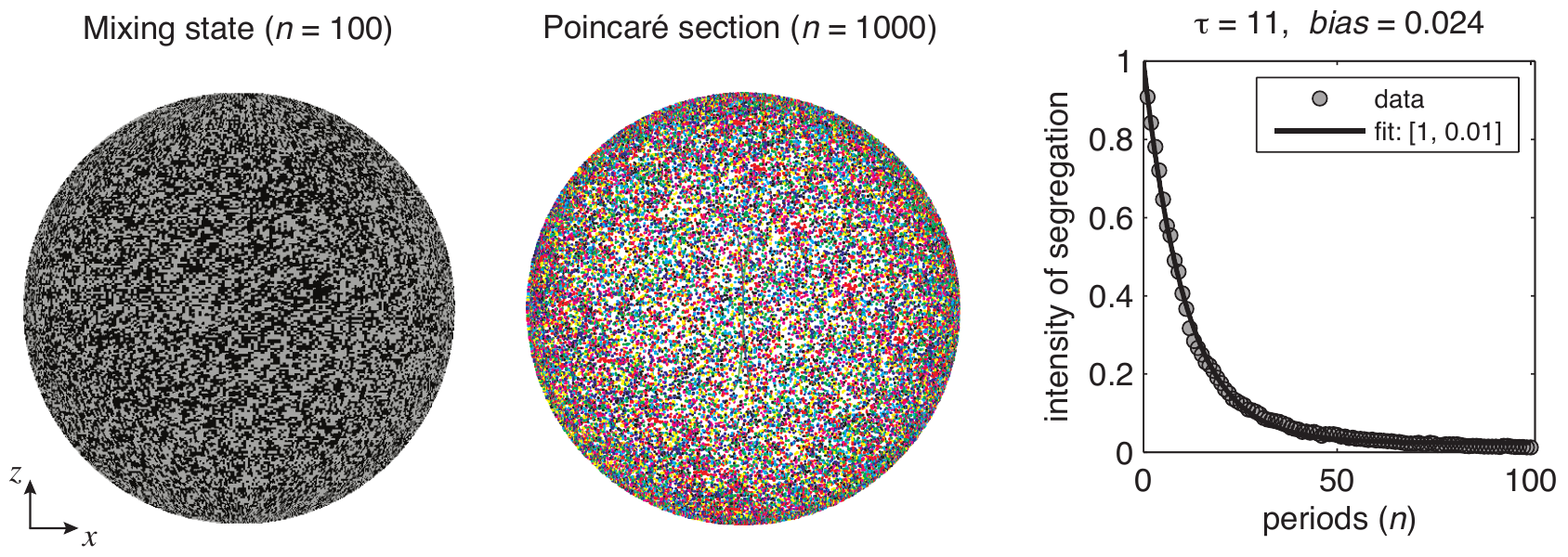}}
\caption{Further examples of mixing behavior in the half-full blinking spherical tumbler with symmetric thin flowing layers: $\epsilon_z = \epsilon_x = 0.05$.}
\label{fig:sym_mix_ex_thin2}
\end{figure}

\begin{figure}
\centering
\subfloat[flowing layer of typical thickness: $\epsilon_z=\epsilon_x = 0.15$]{\includegraphics[width=0.95\textwidth]{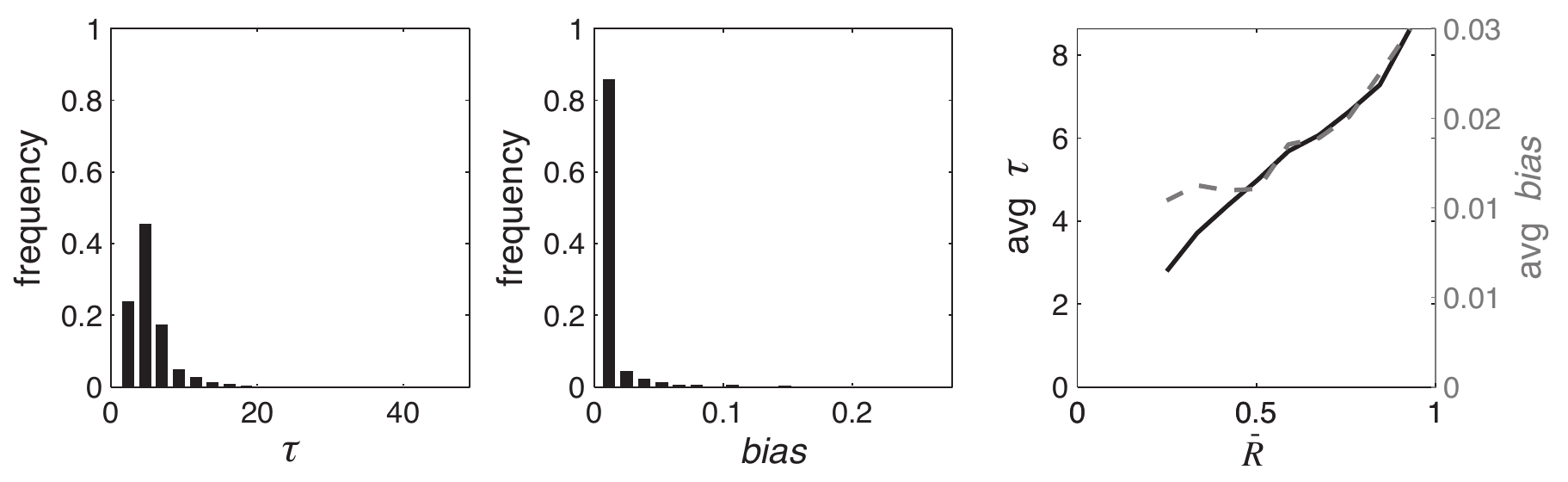}}\\
\subfloat[thin flowing layer: $\epsilon_z=\epsilon_x = 0.05$]{\includegraphics[width=0.95\textwidth]{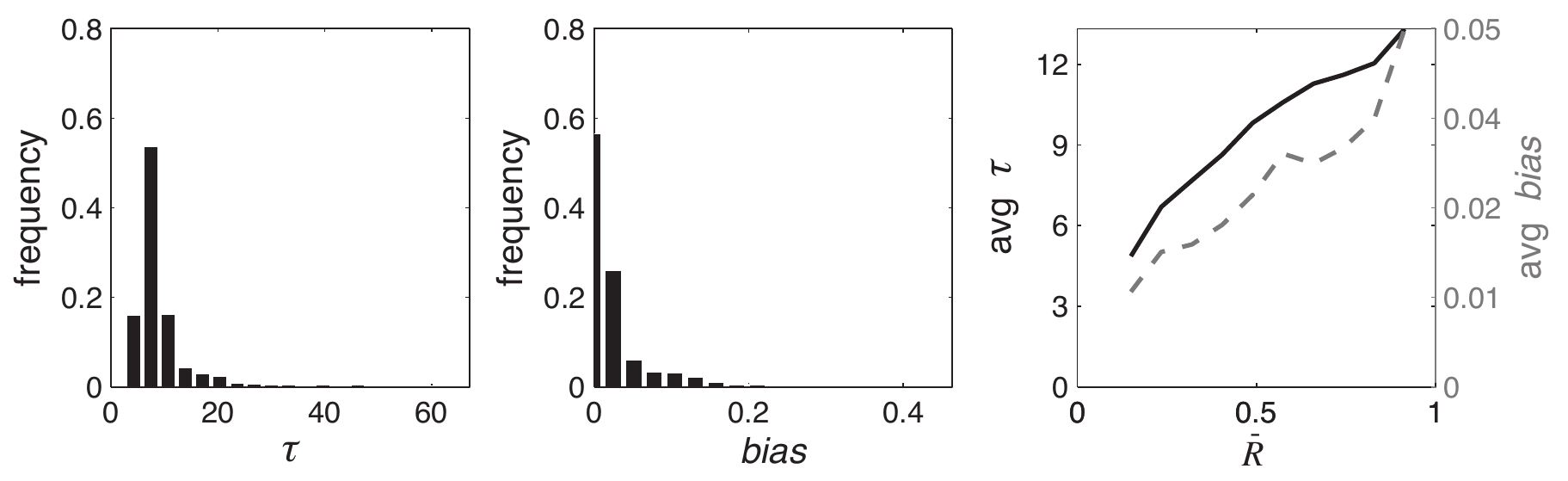}}
\caption{Statistics of the $\tau$-$bias$ scatterplots from Fig.~\ref{fig:tb_scatter}. In the rightmost column, the black solid and dashed gray curves represent the averages of $\tau$ and $bias$, respectively, over all $(\theta_z,\theta_x)$ as functions of $\bar{R}$. For large $\tau$ and large $bias$ there are very few contributions to the histograms, nevertheless we show the full range for completeness. Note the shift of the peak to the bin centered at $bias=0$ in (b).}
\label{fig:tb_scatter_stat}
\end{figure}

To better summarize the data from the scatterplots in Fig.~\ref{fig:tb_scatter}, in Fig.~\ref{fig:tb_scatter_stat} we show some statistics of the distributions of the values of $\tau$ and $bias$. Clearly, the $\tau$ distribution's peak for the case with a thinner flowing layer is shifted to a larger value of $\tau$ showing that mixing is generally worse for a thinner flowing layer. However, for both Fig.~\ref{fig:tb_scatter_stat}(a) and (b) we see that, for most choices of angles, most shells mix quickly and thoroughly with the poor mixing examples falling in the tails of the distributions. Additionally, when all $\tau$ and $bias$ values corresponding to a given $\bar{R}$ are averaged over all choices of $(\theta_z,\theta_x)$ a clear trend emerges. Larger $\bar{R}$ suggest slower and less complete mixing, on average. This can be explained by noting that larger shells have more area, thus mixing is (on average) slower. In addition, the shells with large $\bar{R}$ are those that mix the poorest, by far, when $\theta_z$ and $\theta_x$ are near $5^\circ$, therefore large $\bar{R}$ also correlates with incomplete mixing, on average. One way to motivate this is to realize that for some of the lowest angles of rotation material on the shells with largest radius either never enters the flowing layer or become ``stuck'' in the flowing layer due to the small rotations about each of the two axes.

\section{Conclusion}
The work described here exemplifies the practical difficulty in determining the ``best'' parameters for ``good mixing'' in a physically realistic 3D chaotic flow. Both the time to mix and the degree of mixing are important. The $\tau$-$bias$ approach is useful in identifying which parameters result in good mixing according to both of these measures, but does have some drawbacks. First, it requires methodically testing the entire parameter space (here, $\theta_z$, $\theta_x$, $\epsilon_x=\epsilon_z$ and $\bar{R}$), which is possible due to the availability of a continuum model with an analytic solution \cite{c11,clos15}. Still, studying mixing would be much more difficult with a more complicated flow model or, worse, no model at all. Second, the $\tau$-$bias$ approach only clarifies ``what'' protocols mix or do not mix, but not ``why'' they do or do not. Clarifying the latter still requires interpretation and physical insight. Nevertheless, the results presented here demonstrate a methodology to analyze the quality of mixing in a 3D chaotic system. 
 
\begin{acknowledgement}
I.C.C.\ was supported, in part, by a Walter P. Murphy Fellowship from the Robert R.\ McCormick School of Engineering and Applied Science and by US National Science Foundation grant CMMI-1000469 at Northwestern and by the LANL/LDRD Program through a Feynman Distinguished Fellowship at Los Alamos National Laboratory, which is operated by Los Alamos National Security, L.L.C. for the National Nuclear Security Administration of the U.S. Department of Energy under contract DE-AC52-06NA25396. We thank Stephen Wiggins for suggesting the $\tau$-$bias$ scatterplots and useful discussions.
\end{acknowledgement}
\bibliographystyle{spphys}
\bibliography{granular.bib}
\end{document}